\documentstyle[epsfig,12pt]{article}

\def\beqra{\begin{eqnarray}}
\def\eeqra{\end{eqnarray}}
\def\beq{\begin{equation}}
\def\eeq{\end{equation}}
\def\ds{\displaystyle}

\def\cw{\cos\theta_W}
\def\sw{\sin\theta_W}

\def \lta {\mathrel{\vcenter
     {\hbox{$<$}\nointerlineskip\hbox{$\sim$}}}}
\def \gta {\mathrel{\vcenter
     {\hbox{$>$}\nointerlineskip\hbox{$\sim$}}}}

{

\def\half{\mbox{\small $\frac{1}{2}$}}

\begin{document}

\sloppy

%
%

\title{\bf The Sphaleron in a  Magnetic Field\\
and\\
Electroweak Baryogenesis}
\author{D. Comelli$^1$, D. Grasso$^{2,3}$, M. Pietroni$^3$ and A. Riotto$^4$}
\maketitle
\footnotesize{
\begin{center}
$^1$INFN, Sezione di Ferrara,\\
Via Paradiso 12, I-44100 Ferrara, Italy\\
$^2$Dipartimento di Fisica ``G.Galilei'',\\
Via F. Marzolo 8, I-35131 Padova, Italy\\
$^3$INFN, Sezione di Padova,\\
Via F. Marzolo 8, I-35131 Padova, Italy\\
$^4$ CERN TH-Division, \\
CH-1211, Geneva 23, Switzerland
\end{center}
}

\abstract{The presence of a primordial magnetic field in the early universe  
affects the dynamic of the electroweak phase transition  enhancing its 
strength. This effect may enlarge the window for electroweak baryogenesis 
in the minimal supersymmetric extension of the standard model or even 
 resurrect the electroweak baryogenesis
scenario in the standard model.
We compute the sphaleron energy in the background of the magnetic field
and show that, due to the sphaleron dipole moment, the barrier between
topologically inequivalent vacua is lowered. Therefore, 
the preservation of the 
baryon asymmetry   calls for 
 a much 
stronger phase transition than required in the absence of a magnetic field. 
We show that this effect  overwhelms 
 the gain in the phase transition
strength, and conclude that magnetic fields do not help electroweak 
baryogenesis.}



\vskip0.3in
\noindent
INFNFE--01--99\\DFPD/99/TH/07\\CERN-TH/99-50



\section{Introduction} 
The {\it sphaleron} \cite{KM}, is a static and unstable solution of 
the field equations of the electroweak model, corresponding to the top of
the energy barrier between two topologically distinct vacua. As was first 
noticed by 't Hooft \cite{tH}, topological transitions are associated, via the
axial anomaly, to the violation of baryon (B) and lepton (L) number, 
in the combination B+L. If thermal
fluctuations in the early universe were strong enough to generate sphaleron
configurations, then B+L was badly violated and any previously 
produced B+L--asymmetry would have been washed out. 
In any scenario of baryogenesis 
\cite{RT} it is then crucial to know at which epoch do the sphaleronic
transitions fall out of thermal equilibrium. Generally this happens 
at temperatures below $\bar{T}$ such that \cite{Washout}
\beq
\frac{E(\bar{T})}{\bar{T}} \ge A\,\,,
\label{washout}
\eeq
where $E(T)$ is the sphaleron energy at the temperature $T$ and 
$ A \simeq 35 - 45 $, depending on the poorly known prefactor of the
sphaleron rate.
In the case of baryogenesis at the electroweak scale one requires the
sphalerons to drop out of thermal equilibrium soon after the electroweak 
phase transition. It follows that the requirement $\bar{T}=T_c$, where $T_c$ 
is the critical temperature, turns eq. (\ref{washout}) into a lower bound
on the higgs vacuum expectation value (VEV),
\beq
\frac{v(T_c)}{T_c} \gta 1\,.
\label{first}
\eeq
As a result of intense research in the recent years \cite{EWPT}, 
it is by now agreed that
the standard model (SM) does not have a phase transition strong enough
as to fulfill eq. (\ref{first}), whereas there is still some room left in
the parameter space of the minimal supersymmetric standard model (MSSM) 
\cite{RT}.

In refs.~\cite{GS} an interesting observation was made, which 
could potentially  revive baryogenesis even in the SM; 
if a magnetic field for the hypercharge $U(1)_Y$ was present for $T>T_c$,
the electroweak phase transtion would have been delayed. 
The VEV soon after the transition would then have been larger 
and possibly able to fulfill eq. (\ref{first}). The effect can be understood
in analogy with Meissner effect, {\it i.e.} the expulsion
of the magnetic field from superconductors. In our case, it is the 
$Z$--component of the hypercharge $U(1)_Y$ magnetic field which is expelled
from the broken phase, and this process costs in terms of free energy. Then, 
compared to the case in which no magnetic field is present, it is less
convenient for the system to go to the broken phase, and it has to wait 
until the effective potential becomes so deep as to compensate for the
energy spent to expell the $Z$--component of the field. 

The generation of magnetic fields before -- or during \cite{GraRio} --  the
electroweak phase transition is predicted by several models, as is reviewed
in ref.~\cite{Enqvist} 

The perturbative estimates of refs.~\cite{GS} gave very optimistic indications,
concluding that eq. (\ref{first}) could be fulfilled for higgs masses up to
$m_H \simeq 100$ GeV if a hyper-magnetic field $B_Y \gta 0.3~T^2$ 
existed at the time of the electroweak phase transition. Lattice simulations
\cite{KLPRS} confirmed that the phase transition is 
strengthened by a magnetic field, but not enough as to give
a viable baryogenesis scenario in the SM for $m_H \gta 80$ GeV.

Even if these more pessimistic conclusions were correct, a magnetic field 
could still play a relevant role in models such as the MSSM, enlarging
the portion of the parameter space available for baryogenesis. Analyzing 
further this scenario is therefore   worth while.

The purpose of this letter is to  point out and study an important effect 
which was not taken into account in previous discussions on the role
of magnetic fields in electroweak baryogenesis, namely the large magnetic
dipole moment of the sphaleron \cite{KM}.  
In the background of a magnetic field the coupling with the dipole moment
decreases the height of the sphaleron barrier, so that thermal fluctuations 
are more effective in producing topological transitions. 
In order to fulfill the freeze-out condition, eq.
(\ref{washout}), it is then necessary to have larger VEV's than that of eq. 
(\ref{first}), so that what has been gained in terms of strength of the phase
transition may be lost by the fact that sphalerons have become lighter.
We will show that this is indeed the case, at least for field values which
can be studied reliably with our methods.

The paper is organized as follows. In the next section we shortly review
the main equations which describe the sphaleron for finite values of the 
Weinberg angle. In section 3, an external magnetic field is introduced 
and we will determine its effect on the sphaleron energy and magnetic moment.
Finally, in section 4, we decribe the consequence of our results on
the electroweak baryogenesis scenario.

\section{The Sphaleron dipole moment} 
In the limit of vanishing Weinberg angle, $\theta_w \to 0$, the sphaleron 
is a spherically symmetric, hedgehog-like configuration of $SU(2)$ gauge and 
Higgs fields \cite{FH}.  No magnetic moment is present in this case.
As $\theta_w$ is turned on 
the $U(1)$ field is excited and the spherical symmetry is reduced to an axial 
symmetry \cite{KM,KL,KKB}. The most general Ansatz for the sphaleron at 
$\theta_w \neq 0$ was given in ref.~\cite{KKB}, and requires seven independent
scalar functions of the spherical coordinates $r$ and $\theta$. The 
$\theta$-dependence is however very mild, and a very good approximation to
the exact solution is obtained using the Ansatz by Klinkhamer and Laterveer
\cite{KL}, which requires four scalar functions of $r$ only, 
\beqra
&&\ds g^\prime a_i \, dx^i = (1 - f_0(\xi))\, F_3 \,,\nonumber\\
&&\ds g W^a_i \sigma^a \, dx^i = (1-f(\xi)) (F_1 \sigma^1 +F_2 \sigma^2) +
(1-f_3(\xi)) F_3 \sigma^3 \,,\nonumber\\
&&{\bf \Phi} = \frac{v}{\sqrt{2}} \left(
\begin{array}{c}
\ds 0 \\
\ds h(\xi) 
\end{array}
\right)\,\,, 
\label{ansatz}
\eeqra
where $g$ and $g^\prime$ are the $SU(2)_L$ and $U(1)_Y$ gauge couplings, 
$v$ is the higgs VEV such that $M_W= g v/2$, $M_h = \sqrt{2 \lambda} v$, 
$\xi=gvr$, $\sigma^a$ ($a= 1,2,3$) are the Pauli matrices, and the $F_a$'s 
are 1-forms defined as follows \cite{KL}
\beqra
&&\ds 
F_1 = - 2 \sin\phi\,d\theta - \sin 2 \theta \cos \phi \,d\phi\,,\nonumber\\ 
&&\ds 
F_2 = - 2 \cos\phi\, d\theta + \sin 2 \theta \sin \phi \,d\phi\,,\nonumber\\
&& \ds
F_3 =\,\,\,\, 2 \sin^2 \theta \,d\phi \,.\nonumber
\eeqra 

The boundary conditions for the four scalar functions are 
\[
f(\xi)\,,\,f_3(\xi)\,,\,h(\xi) \to 0\,,\,\,\,\;\;\,f_0(\xi)\to 1\;\;
\;\;\;\,\,\,\,\,\,\,\,\,\,\,\,\;\;\;\;\;\;\;\;\;\;\;\;\;\;\;\;
{\mathrm for}\,\;\, \xi\to 0
\]
\beq
\,\;\;\;f(\xi)\,,\,f_3(\xi)\,,\,h(\xi)\,,\,f_0(\xi) \to 1\,,\,\,\,\;\;\,\;\;
\;\;\;\,\,\,\,\,\,\,\,\,\,\;\;\;\;\;\;\;\;\;\;\;\;\;\;\:\;\;\;\;\;\;\;
{\mathrm for}\,\;\, \xi\to \infty\,.
\label{bcinf}
\eeq

In ref.~\cite{KKB} it was shown that the above Ansatz can be recovered at
the first order in a Legendre polynomials expansion of
the exact sphaleron solution at $\theta_w \neq 0$, giving an excellent  
approximation to the latter for $\theta_w$ of the order of the physical value 
$\theta_w=0.5$ rads.
Taking the $\theta_w \to 0$ limit the field equations give 
\[
f_3(\xi) \to f(\xi)\,,\;\;\;\;\;\; \,f_0(\xi)\to 1\;\;
\;\;\;\,\,\,\,\,\,\,\,\,\,\,\,\;\;\;\;\;\;\;\;\;\;\;\;\;\;\;\;
({\mathrm for}\,\;\, \theta_w\to 0)
\] 
for any value of $\xi$, thus reproducing the two-function Ansatz of the pure 
$SU(2)$-Higgs case \cite{KM}.
The source of the hypercharge field $a_i$ is the $O(\theta_w)$ current $J_i$
\[
\partial_i f_{ij} =   J_i 
\]
where  
\beqra
&& \ds f_{ij}= \partial_i a_j -\partial_j a_i\,,\nonumber\\
&&\ds J_i  = - \frac{1}{2} ig^\prime \left[ {\bf \Phi}^\dagger D_i  
{\bf \Phi} - (D_i {\bf \Phi})^\dagger {\bf \Phi} \right] \nonumber\,,\\
&& \ds D_i {\bf \Phi} = \partial_i {\bf \Phi} - \half i g \sigma^a W^a_i 
{\bf \Phi} - \half i g^\prime a_i {\bf \Phi}\;.\nonumber 
\eeqra
The hypercharge field behaves asymptotically as a pure dipole
\[
a_i \to  \epsilon_{ijk} \frac{\mu_j \, x_k}{4 \pi r^3}\,,
\]
with the dipole moment $\mu_j = \delta_{j3}\, \mu$.
At the first order in $\theta_w$, $a_i$ can be neglected in $J_i$, which then 
takes the form
\beq
J_i^{(1)} = -  \half g^\prime v^2 \frac{h^2(\xi) 
[1 - f(\xi)]}{r^2}\, \epsilon_{3ij} x_j
\;,
\label{cur_0}
\eeq
where $h$ and $f$ are the solutions in the $\theta_w \to 0$ limit, giving for 
the dipole moment \cite{KM}
\beq
\mu^{(1)} = \frac{2 \pi}{3}  \frac{g^\prime}{g^3 v} \int_0^\infty d\xi
\xi^2 h^2(\xi) [1-f(\xi)]\;.
\label{mu1}
\eeq

The reader should note that the dipole moment is a true electromagnetic one 
 because in the broken phase only the electromagnetic component of
the hypercharge field survives at long distances.

\section{Sphaleron in a magnetic field} 
If an external hypercharge field, $a^c_i$, is turned on, 
the energy functional is modified as
\beq 
E = E_0 - E_{\mathrm dip}\,,
\label{energy}
\eeq
with
\beqra
&& \ds  E_0= \int d^3x \left[ \frac{1}{4} F^a_{ij}F^a_{ij} + 
\frac{1}{4} f_{ij}f_{ij} + (D_i {\bf \Phi})^\dagger (D_i {\bf \Phi}) +
V({\bf \Phi}) \right]\;,\nonumber\\
&& \ds F^a_{ij} = \partial_i W^a_j - \partial_j W^a_i +g \epsilon^{abc} W^b_i
W^c_j\;,\nonumber\\
&&\ds V({\bf \Phi}) = \lambda \left( {\bf \Phi^\dagger} {\bf \Phi} -
\half v^2\right)^2\;,\nonumber
\eeqra
and
\beq
\label{intenergy}
E_{\mathrm dip} = \int d^3x J_i a^c_i = \half \int d^3x f_{ij}f^c_{ij} 
\eeq

Besides the SM, our treatment will also  cover the case of the 
 MSSM in the parameter space in which
one of the two neutral higgs directions decouples, {\it i.e.} when the 
mass of the pseudoscalar $A^0$ is much larger than the $Z$-boson mass and
the low-energy potential reduces to a one-dimensional SM-like
potential. Even though this is not the most favorable situation 
for  the generation
of the baryon asymmetry -- since the latter turns out to be proportional
to the variation of the ratio between the 
VEV's of the two neutral Higgs fields inside the bubble wall \cite{var} --
the strength of the MSSM phase transition is stronger. Our 
 approximation is therefore  well justified here, 
and eq. (\ref{energy}) can be 
thought as the  energy for this effective theory even when
thermal corrections are included and the VEV of the Higgs field $v$ should
be thought as temperature dependent.

We will consider a constant external hypermagnetic field $B^c_Y$ directed 
along the $x_3$ axis, {\it i.e.}
\beq 
a_i^c = - \epsilon_{3ij} \frac{B^c_Y}{2} x_j\,.
\label{Bc}
\eeq 
In the $\theta_w\to0$ limit the sphaleron has no hypercharge contribution and
then $E_{\mathrm dip}^{(0)} = 0 $. 
At $O(\theta_w)$, using (\ref{cur_0}) and (\ref{Bc}) we get a 
simple dipole interaction 
\beq
E_{\mathrm dip}^{(1)} = \mu^{(1)} B^c_Y \; .
\label{lino}
\eeq
When the external field $B^c_Y$ is zero, the $O(\theta_w)$ approximation
is known to give values of the sphaleron energy with a better than 
 percent accuracy \cite{KL,KKB}. 
On the other hand, comparing $E_{\mathrm dip}^{(1)}$ with $E_0$ 
-- which is typically $O(M_W/g)$ -- we  realize 
that when $B^c_Y\neq 0$ the expansion parameter is effectively 
$\theta_w B^c_Y/(g v^2)$
so that, for large values of $B^c_Y$, higher orders in $\theta_w$ 
may become important. In order to assess the range of validity of the 
approximation (\ref{lino}) we need to go beyond the leading order in 
$\theta_w$ and look for a nonlinear $B^c_Y$--dependence of $E$.

First of all, we notice that the external field (\ref{Bc})
does not break the axial symmetry, so we will use the same Ansatz as for the 
$ B^c_Y=0$ case, eq. (\ref{ansatz}). Moreover, the field equations are
left unchanged by a constant $B^c_Y$ ($\partial_i f^c_{ij} =0$), {\it i.e.}
\beqra
&& \ds f''+
\frac{1-f}{4\, \xi^2}
\left[ 8(f(f-2)+f_3+f_3^2)+\xi^2 h^2 \right]=0\,, \nonumber\\
&& \ds f_3''-\frac{2}{ \xi^2}\left[3 f_3+ f (f-2)
(1+2 f_3)\right]-  \frac{h^2}{4} (f_3-f_0)=0\,,\nonumber\\
 &&
\ds f_0''+\frac{g_1^2}{4\, g^2}\, h^2 (f_3-f_0)+2\,
\frac{1-f_0}{\xi^2 }=0\,,\nonumber \\
&&\ds 
h''+\frac{2}{\xi} h' 
-\frac{2}{3 \xi^2}\,
 h \left[2(f-1)^2 +(f_3-f_0)^2\right] -\frac{2 \lambda }{3\,g^2} (h^2 - 1) h
=0\,. \label{system}
\eeqra
The only modification induced by $B^c_Y$ resides in the boundary conditions
since -- as $r \to \infty$ -- we now have 
\beqra
\label{bcinfi}
&& a_i = \cw A_i - \sw Z_i \to \cw A^c_i = \cos^2\theta_w a^c_i \nonumber \\
&& W^3_i = \sw A_i + \cw Z_i \to \sw A^c_i = \cw\sw a^c_i \nonumber \,.\\
\eeqra
The equalities on the right side of (\ref{bcinfi}) have been obtained by 
requiring the continuity of the gauge fields at the boundary between
the broken and the symmetric phase \cite{GS}. 
As a consequence of the new boundary conditions (\ref{bcinfi}), 
eqs.(\ref{bcinf}) are turned into
\[ f(\xi)\,,h(\xi) \to 1\,,\;\;\;\;\;\;\;\; f_3(\xi)\,,f_0(\xi) \to 1-B^c_Y 
\sin 2\theta_w \frac{\xi^2}{8 g v^2}\,\;\;\;\;\;\;\;\;\;\;\;
({\mathrm for} \;\xi \to \infty)\;
\]
wheras the boundary condition for $\xi \to 0$ are left unchanged.
In order to keep the same form of the boundary conditions as for the 
case without the external field, it is convenient to define two new scalar
functions, $g_0(\xi)$ and $g_3(\xi)$, as
\beq
g_{0(3)}(\xi) \equiv f_{0(3)}(\xi)
+ B^c_Y \sin 2\theta_w \frac{\xi^2}{8 g v^2}\;.
\label{sosti}
\eeq 
The functions $f(\xi),~g_0(\xi),~g_3(\xi)$ and  $h(\xi)$
have been determined by solving numerically eqs.(\ref{system}).
The result is plotted in fig.1 for $B^c_Y=0$ (solid lines) and 
$B^c_Y = 0.19~v^2$ (dashed). The modifications between the two groups of 
curves is responsible for the nonlinear dependence of $E$ on $B^c_Y$ 
mentioned above.

\begin{figure}[t]
\centerline{\epsfxsize=4.5in\epsfbox{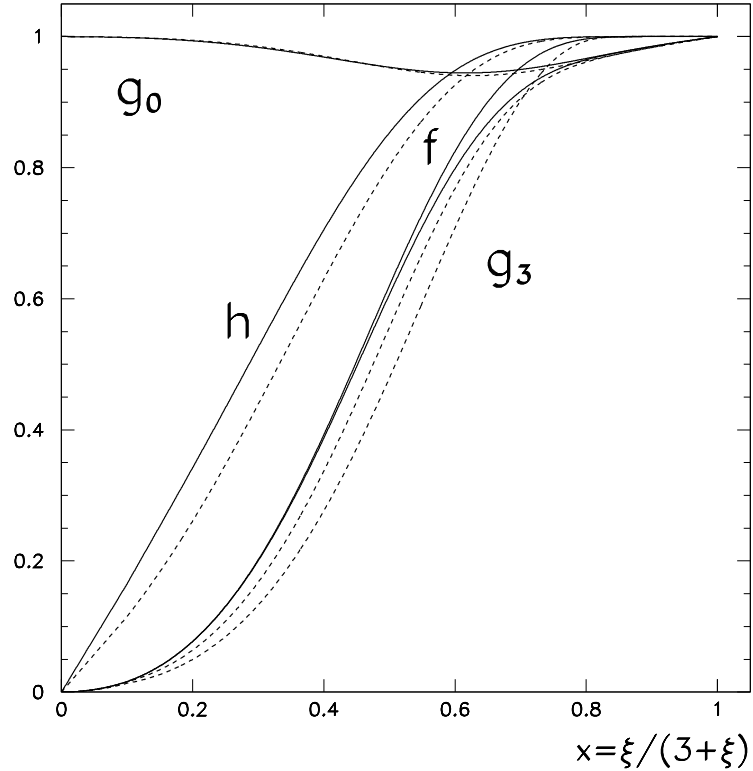}}
\caption{The scalar functions of the Ansatz (\ref{ansatz}) --
 redefined according to (\ref{sosti}) -- for $M_h=M_W$ and $B^c_Y = 0$ (solid),
$0.2~v^2$ (dashed).
}
\end{figure}

In tab. 1 the values of $E$ and of the effective electromagnetic dipole moment 
$\mu_{eff}$,
defined as
\beq
\mu_{eff}(B^c_Y)=\frac{\Delta E}{\cw B^c_Y}\;,
\label{linear}
\eeq
whith $\Delta E \equiv E(B^c_Y) - E(B^c_Y=0)$, are given for $M_h=M_W$. 
In the $B^c_Y \to 0$ limit, $\mu_{eff}$ tends to $\mu^{(1)}$ in eq. 
(\ref{mu1}).
From the variation of $\mu_{eff}$ in the considered $B^c_Y$--range we 
conclude that the corrections to the linear approximation
\[
\Delta E \simeq \mu^{(1)} \cw B^c_Y
\] 
are less than $5 \%$.

For larger values of $B^c_Y$ non-linear effects increase sharply. However,
for such large magnetic fields the broken phase of the SM is believed to become
unstable to the formation either of $W$-condensates \cite{AO} or of
 a mixed phase \cite{KLPRS}. In such situations
the sphaleron solution  does not exist any more, 
so we will limit our considerations to safe values $B^c_Y \lta 0.4~T^2$.

\begin{table}[htb]
{\centering
\begin{tabular}{||c|c|c||}
\hline
\hline
$B^c_Y/v^2$ & $  \varepsilon $ &$m_{eff}$ \\
\hline
$0$ &$ 1.81$  & $-$   \\
\hline
$6 \,10^{-3}$ &$1.79 $  & $1.81$    \\
\hline
$3 \,10^{-2} $ &$1.73$  & $ 1.82$ \\
\hline
$6 \,10^{-2} $ &$ 1.66$  & $ 1.82$     \\
\hline
$0.12$ &$1.51$ &$1.86$  \\
\hline
$0.25$ &$1.20$ &$1.88$  \\
\hline
\hline
\end{tabular}
\par}
\vskip 0.5 cm
\caption{ Values for $\varepsilon=E\,\frac{g}{4\, \pi\,v}$ and 
$ m_{eff} = \mu_{eff} \, \frac{ \alpha_W \, M_W}{e} $ (see text) 
obtained for $M_h=M_W$.}
\end{table}


\section{Consequences for EW baryogenesis}
We are now ready to discuss the effect of the 
sphaleron--magnetic field  dipole interaction on the baryogenesis scenario 
discussed in refs.~\cite{GS}. Let us first briefly recall its salient feature,
namely the possibility of getting a stronger phase transition when the 
symmetric phase is permeated by a hypermagnetic field. 
Indeed, when this is the case, the pressure in the symmetric phase gets an 
extra contribution
\[
P_u = -V(0) + \half {B^c_Y}^2\,.
\]
In the broken phase only the electromagnetic component of the hypercharge
field survives at long distances, the $Z$--component being screened as the
$Z$--boson becomes massive, thus
\[
P_b = -V(\Phi) + \half \cw^2 {B^c_Y}^2\,.
\]
The phase transition takes place at the temperature $T_c$ where
phase equilibration  occurs, 
\beq
V(0)-V(\phi) = \half \sin^2\theta_w {B^c_Y}^2\,,\;\;\;\;\;\;\;\;\;\;\;\;\;\;
({\mathrm for}\;\;\;T=T_c)\,.
\label{equil}
\eeq
Considering the simple approximation to the SM potential,
\[ V(\Phi) \simeq a (T^2 - T_b^2) \Phi^2 +\frac{\lambda}{4} \Phi^4\,,\]
where the constant $a$ needs not to be specified here, eq.(\ref{equil}) 
gives
\beq
\frac{v^2(T_c)}{T_c^2} = b \:\sqrt{\frac{2 \sin^2\theta_w}{\lambda}} \,,
\label{si?}
\eeq
where $b \equiv B^c_Y /T_c^2$.
If we fix $M_h=M_W$, we get $v(T_c)/T_c \gta 1$ for 
${B^c_Y}\gta 0.33~{T_c^2}$. The phase transition is then of the 
first order even in absence of a cubic term in the effective potential 
and the bound (\ref{first}) is satisfied.

The conclusion that the sphaleron freeze-out condition (\ref{washout}) is 
satisfied and the baryon asymmetry preserved is however premature. Indeed,
in an external magnetic field the relation between the VEV and the sphaleron
energy is altered and eq. (\ref{first}) does not imply (\ref{washout})
any more. We can understand it by considering the linear approximation 
to $E$, 
\beq
E\simeq E(B^c_Y=0) - \mu^{(1)} B^c_Y \cw \equiv \frac{4 \pi v}{g} 
\left(\varepsilon_0 - \frac{\sin 2\theta_w}{g} \frac{B^c_Y}{v^2} m^{(1)}
\right) 
\label{newwash}
\eeq
where $m^{(1)}$ is the $O(\theta_W)$ dipole moment expressed in 
units of $e/\alpha_W M_W(T)$.

If $B^c_Y=0$ the freeze-out condition $E/T>A$ translates straightforwardly
into \mbox{$v(T_c)/T_c > x_0 \equiv A g/ 4 \pi \varepsilon_0$},
which -- again for $M_h=M_W$
and $A=35$ -- gives (see tab.~1) $x_0 \simeq 1$.
As $B^c_Y$ is turned on, eq. (\ref{newwash}) gives
\beq
\frac{v(T_c)}{T_c} > \frac{x_0}{2}\left(1+\sqrt{1+
\frac{4 \sin 2\theta_w  \, m^{(1)}\, b}{g \,\varepsilon_0 \,x_0^2}}\right)\,,
\label{no}
\eeq
which gives $v /T_c \gta 1.3$ for the same parameters as above and $b=0.33$.
Using (\ref{si?}) in (\ref{no}) we see that there is no value of the field $b$
for which the gain in terms
of phase transition strength is enough to push the sphaleronic transitions 
out of thermal equilibrium.

The above estimate is confirmed by more accurate computations, as plotted
in fig. 2, where our numerical values for $E$ and the one-loop effective
potential were used. 
\begin{figure}[t]
\centerline{\epsfxsize=4.5in\epsfbox{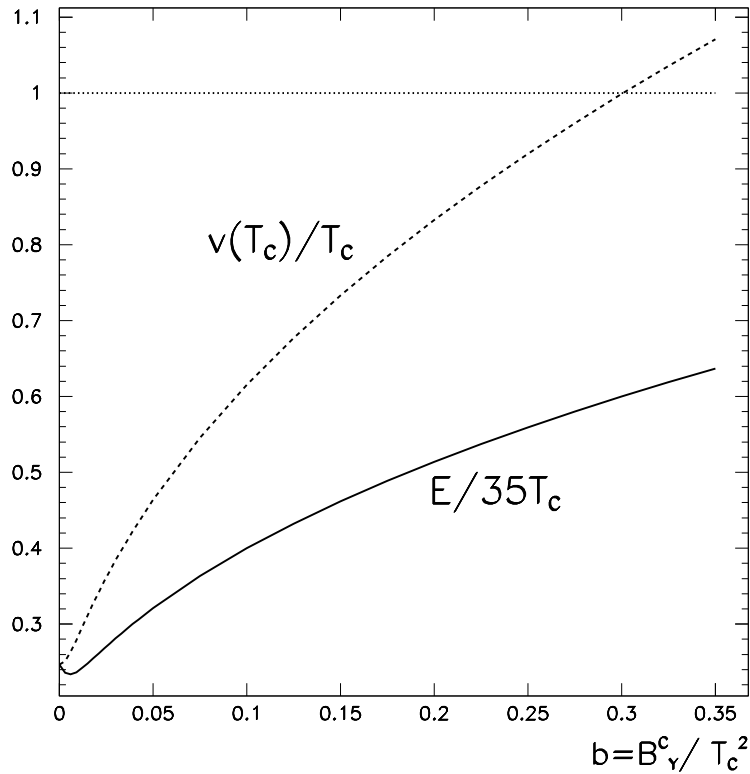}}
\caption{The VEV at the critical temperature, $v(T_c)$, and the 
sphaleron
energy {\em vs.} the external magnetic field for
$M_h=M_W$. Even if $v(T_c)/T_c \gta 1$ the washout condition $E/T_c \gta 35$
is far from being fulfilled.
}
\end{figure}

\section{Summary}
In this letter we  have computed the sphaleron energy in the background 
of a constant magnetic field. 
Our main motivation was the investigation of a recently proposed scenario of 
electroweak baryogenesis, in which the phase transition is strengthened
by a background hypermagnetic field.
We have confirmed this effect, but pointed out that it is not enough as to
preserve the baryon asymmetry produced before or during the phase transition.
The main point is the dipole interaction between the sphaleron and the magnetic
field which lowers the energy barrier between topologically inequivalent
vacua, thus requiring a phase transition much stronger than that obtained
by perturbative computations.


\end{document}